\documentclass[aps,prb,twocolumn,groupedaddress]{revtex4-1}
\usepackage[latin9]{inputenc}
\usepackage{graphicx}
\usepackage{epstopdf}
\makeatletter
 
 \@ifundefined{textcolor}{}
 {%
   \definecolor{BLACK}{gray}{0}
   \definecolor{WHITE}{gray}{1}
   \definecolor{RED}{rgb}{1,0,0}
   \definecolor{GREEN}{rgb}{0,1,0}
   \definecolor{BLUE}{rgb}{0,0,1}
   \definecolor{CYAN}{cmyk}{1,0,0,0}
   \definecolor{MAGENTA}{cmyk}{0,1,0,0}
   \definecolor{YELLOW}{cmyk}{0,0,1,0}
 }

\usepackage{dcolumn}
\usepackage{bm}

\makeatother

\begin{document}






\title{Low-frequency phonons of few-layer graphene within a tight-binding model}

\author{Valentin N. Popov}

\affiliation{Faculty of Physics, University of Sofia, BG-1164 Sofia, Bulgaria}

\author{Christian Van Alsenoy}

\affiliation{University of Antwerp, Chemistry Department, Groenenborgerlaan 171, B-2020 Antwerp, Belgium}

\date{\today}
\begin{abstract}
Few-layer graphene is a layered carbon material with covalent bonding in the layers and weak van der Waals interactions between the layers. The interlayer energy is more than two orders of magnitude smaller than the intralayer one, which hinders the description of the static and dynamic properties within electron band structure models. We overcome this difficulty by introducing two sets of matrix elements - one set for the covalent bonds in the graphene layers and another one for the van der Waals interactions between adjacent graphene layers in a tight-binding model of the band structure. Both sets of matrix elements are derived from an ab-initio study on carbon dimers. The matrix elements are applied in the calculation of the phonon dispersion of graphite and few-layer graphene with AB and ABC layer stacking. The results for few-layer graphene with AB stacking agree well with the available experimental data, which justifies the application of the matrix elements to other layered carbon structures with van der Waals interactions such as few-layer graphene nanoribbons, multiwall carbon nanotubes, and carbon onions.      
\end{abstract}
\maketitle


\section{Introduction}

Few-layer graphene (FLG) is a new material with extraordinary properties and plenty of  prospective applications.\cite{geim09} It consists of parallel graphene layers bound together by weak van der Waals (vdW) interactions. Although these interactions are much weaker than the covalent bonds between the carbon atoms within the layers, they give rise to measurable changes in the electronic band structure and phonon dispersion of FLG.\cite{nguy14} In particular, the low-frequency Raman spectra in the region below about $300$ cm$^{-1}$ exhibit distinct features due to resonant scattering from quasi-acoustic phonons with shearing and breathing atomic motion of the adjacent layers, so-called shearing modes (SMs) and breathing modes (BMs).\cite{tan12,lui12,cong14} The observed position and shape of these Raman features can be used for characterization of FLG, supported by theoretical predictions.

The weak vdW interactions are usually described by interatomic pair potentials with parameters fitted to the experimental  interlayer separation and elastic constant $C_{33}$ of graphite.\cite{giri56,lii89} Such empirical potentials overestimate the cohesive energy compared to the experimental one. The ab-initio calculations within the density function theory (DFT) in the local density approximation (LDA) reproduce well the measured interlayer separation, while those within the DFT in the generalized gradient approximation (GGA) overestimate this quantity.\cite{moun05} The former approximation predicts almost twice smaller cohesive energy with respect to the experimental ones and the latter one yields even lower  cohesive energy.\cite{hase04}

The weakness of the interlayer vdW interactions, compared to the intralayer covalent bonds, poses a serious problem before the modeling of the quasi-acoustic motion of the layers of graphite and FLG. The interlayer interaction has little effect on the high-frequency phonons in graphite but is crucial for the quasi-acoustic phonon branches. The force-constant lattice-dynamical models with many adjustable parameters fitted to the experimental interlayer separation and quasi-acoustic phonon branches of graphite render a very good description of these branches both in graphite and FLG.\cite{mich08,sato11} The dynamical models utilizing the Lennard-Jones potential,\cite{giri56} Rahaman-Stillinger-Lemberg potential,\cite{chen13} Tersoff-Brenner Bond Order potential,\cite{bren90} and Long-range Carbon Bond Order potential\cite{los05} generally work well for the BM but perform poorly for the SM.\cite{kars11} The DFT approach with dynamical matrix derived from perturbation theory (DFPT) predicts with moderately good accuracy the quasi-acoustic branches of graphite.\cite{pavo96,wirt04} However, the same approach is not as successful for FLG, usually yielding SMs and BMs deviating by up to several per cent from the observed ones.\cite{saha08,yan08,tan12} The force constant models suffer the drawback of not being transferable to other layered carbon structures. The models using empirical potentials or the ab-initio approach have not so far demonstrated acceptable accuracy for the SM and BM. Moreover, the ab-initio  calculations are becoming rather time and resource consuming with the increase of the system size. We show that these drawbacks of the empirical models and ab-initio approach can be overcome within a non-orthogonal tight-binding model with matrix elements, derived from a DFT study on carbon dimers. This model has so far been applied to covalently bonded systems\cite{pore95} but not to systems with vdW interactions. 

Here, we report the calculation of the phonon dispersion of graphite and FLG with AB and ABC layer stacking within the non-orthogonal tight-binding model with matrix elements, derived within the DFT approach. The difference between the  total energy and the band energy of the system is modeled by a pair potential with parameters fitted to the interlayer separation, SM and BM of graphite. The paper is organized as follows. The computational details are presented in Sec. II. The accomplished work is given and discussed in Sec. III, namely, the derivation of the interlayer tight-binding matrix elements (Subsec. III.A), the pair potential (Subsec. III.B), the cohesive energy of FLG (Subsec. III.C), the low-frequency phonon branches of FLG, and the corresponding $\Gamma$-point phonons (Subsec. III.D). The paper ends up with a conclusion, Sec. IV. 

\section{Theoretical background}

The calculation of the phonon dispersion of FLG is performed within a non-orthogonal tight-binding model with two different sets of tight-binding matrix elements for the description of the intralayer and interlayer interactions. The intralayer matrix elements are taken from Ref.~\onlinecite{pore95} and the interlayer ones are derived here using the computational scheme of the latter reference. Namely, the wavefunctions of the electronic system are expanded in terms of localized atomic orbitals with orbital momentum quantum number $0$ and $1$, so-called $s$ and $p$ orbitals. Within the DFT-LDA, the Kohn-Sham equation is solved for carbon dimers to determine the matrix elements of the Hamiltinian $H$ and the overlap matrix elements $S$ between two types of orbitals located at the two carbon atoms of the dimer. For example, for atoms along the $x$ axis, the matrix elements between the pairs of orbitals ($s$,$s$), ($s$,$p_x$), ($p_x$,$p_x$), and ($p_y$,$p_y$) are $S_{ss\sigma}$, $S_{sp\sigma}$, $S_{pp\sigma}$, and $S_{pp\pi}$, respectively. The Slater-Koster scheme is used to express the angular dependence of the matrix elements for differently oriented carbon pairs in the investigated carbon structure. In order to obtain usable matrix elements, the atomic potential is modified by adding the term $(r/r_0)^2$, where $r$ is the electron radial coordinate and $r_0$ is taken equal to twice the covalent atomic radius of carbon ($r_0 = 1.42$ \AA). The resulting electron density is compressed compared to that of the free atom. The derived matrix elements were shown to describe well the electron levels of carbon clusters.\cite{pore95} The total electronic energy is presented as the sum of the band energy, computed from the occupied electron bands, and an additional energy, accounting for the remaining electron correlations. The additional energy is represented as the sum of pair potentials. The obtained matrix elements and pair potentials for the intralayer electrons have been shown to perform well in the calculation of the electron and phonon dispersion of carbon nanotubes\cite{popo10a} and graphene.\cite{popo10}

We extend this approach to derive the interlayer matrix elements and pair potential and apply them to the study of the static and dynamic properties of graphite and FLG with Bernal (AB) and Rhombohedral (ABC) stacking of the layers (Fig. 1). FLG with $N$ layers with AB and ABC stacking will be denoted as NAB and NABC, respectively. For example, 4AB stands for ABAB and 5ABC stands for ABCAB.  

\begin{figure}[tbph]
\includegraphics[width=80mm]{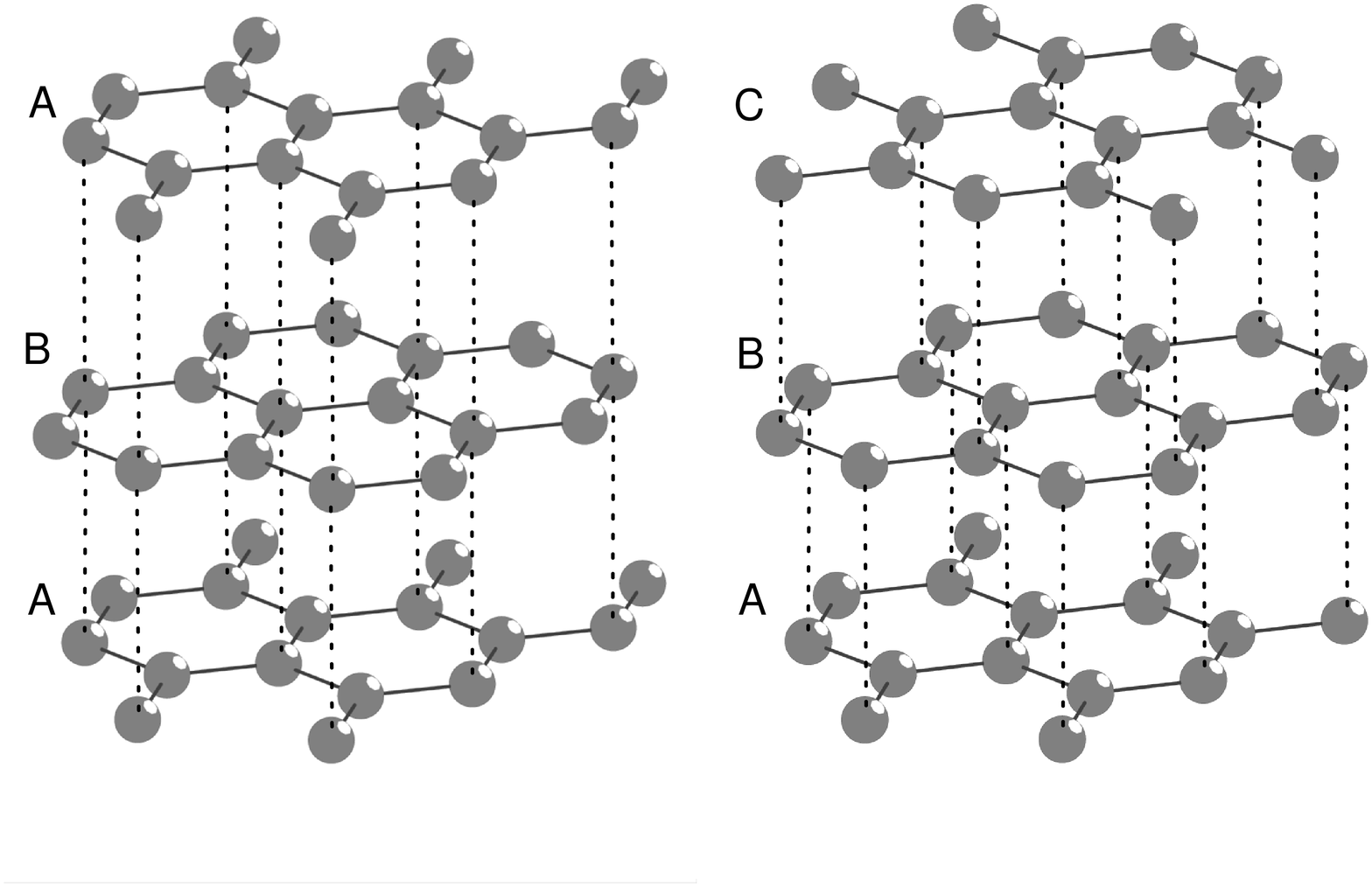} 
\caption{Schematic representation of AB and ABC stacking of graphene layers.}
\end{figure}

\section{Results and Discussion}

\subsection{Interlayer matrix elements}

\begin{table*}[t]
\caption{\label{tab:table1}
Coefficients of the expansion of the interlayer matrix elements of the Hamiltonian $H$ (in a.u.) and the overlap matrix elements $S$ in Chebyshev polynomials $T_m$ in the form: $f(r)=
\sum_{m=1}^{10} c_mT_{m-1}(y) - c_1/2$, where $r$ is the interatomic separation and $y=\left(2r-\left(b+a\right)\right)/\left(b-a\right)$. The boundaries for the polynomials $\left(a,b \right)$ are $\left(1,10 \right)$ (in a.u.).}
\begin{ruledtabular}
\begin{tabular}{lrrrrrrrrr}
&$H_{ss\sigma}$&$H_{sp\sigma}$&$H_{pp\sigma}$&$H_{pp\pi}$&$S_{ss\sigma}$&$S_{sp\sigma}$&$S_{pp\sigma}$&$S_{pp\pi}$\\
\hline
$c_1$&-0.5286482 &  0.3865122 &  0.1727212 & -0.3969243 &  0.4524096 & -0.3509680 & -0.0571487 &  0.3797305 & \\	
$c_2$&0.4368816 & -0.2909735 & -0.0937225 &  0.3477657 & -0.3678693 &  0.2526017 & -0.0291832 & -0.3199876 & \\	
$c_3$&-0.2390807 &  0.1005869 & -0.0445544 & -0.2357499 &  0.1903822 & -0.0661301 &  0.1558650 &  0.1897988 & \\	
$c_4$&0.0701587 &  0.0340820 &  0.1114266 &  0.1257478 & -0.0484968 &	 -0.0465212 & -0.1665997 & -0.0754124 & \\	
$c_5$&0.0106355 & -0.0705311 & -0.0978079 & -0.0535682 & -0.0099673 &  0.0572892 &  0.0921727 &  0.0156376 & \\	
$c_6$&-0.0258943 &  0.0528565 &  0.0577363 &  0.0181983 &  0.0153765 &	 -0.0289944 & -0.0268106 &  0.0025976 & \\	
$c_7$& 0.0169584 & -0.0270332 & -0.0262833 & -0.0046855 & -0.0071442 &  0.0078424 &  0.0002240 & -0.0039498 & \\	
$c_8$&-0.0070929 &  0.0103844 &  0.0094388 &  0.0007303 &  0.0017435 &	 -0.0004892 &  0.0040319 &  0.0020581 & \\	
$c_9$&0.0019797 & -0.0028724 & -0.0024695 &  0.0000225 & -0.0001224 & -0.0004677 & -0.0022450 & -0.0007114 & \\	
$c_{10}$&-0.0003034 &  0.0004584 &  0.0003863 & -0.0000393 & -0.0000443 &  0.0001590 &  0.0005596 &  0.0001427 & \\	
\end{tabular}
\end{ruledtabular}
\end{table*}

We use the computational scheme of Ref.~\onlinecite{pore95} to derive the interlayer matrix elements for graphite and FLG. We use $6-31$G Gaussian basis set for expanding the electron wavefunction and the Vosko-Wilk-Nusair correlation functional in the DFT-LDA  calculations. The parameter $r_0$ is taken equal to the measured interlayer separation in graphite ($r_0 = 3.35$ \AA).\cite{dono82} The interlayer  matrix elements are  shown as a function of the interatomic separation in comparison with the intralayer ones in Figs. 2 and 3, and the coefficients of their expansion in Chebyshev polynomials are given in Table I. The matrix elements are used for calculation of the electron band structure. The total band energy is expressed as the sum of the band energy and an additional energy, expressed in terms of pair potentials.

The phonon calculations are performed with dynamical matrix, derived in quantum-mechanical perturbation theory with electron/hole-phonon matrix element and electron/hole lifetime, derived within the tight-binding model.\cite{popo13a} In particular, the total energy of the distorted crystal lattice is expanded in power series in the atomic displacements up to second order. The series expansion consists of terms with second-order change of the matrix elements in the displacements treated in first-order perturbation theory and terms with first-order change of the matrix elements in the displacements treated in second-order perturbation theory. The electron band structure and phonon dispersion calculations require summation over the Brillouin zone, which is performed over a Monkhorst-Pack mesh of k points of size $40\times 40$ and $40\times 40 \times 4$ for FLG and graphite, respectively. The lattice parameter and phonon frequencies of the studied structures are converged to better than $0.01$\AA~ and  $1$ cm$^{-1}$, respectively.

\begin{figure}[tbph]
\includegraphics[width=80mm]{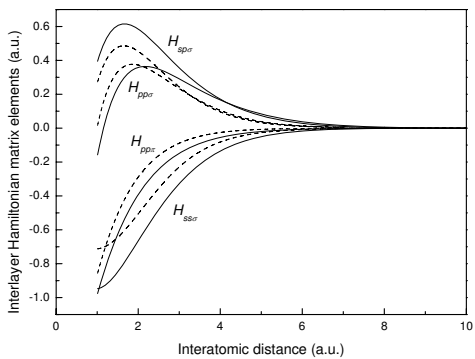} 
\caption{Interlayer matrix elements of the Hamiltonian $H$ (solid lines) as a function of interatomic distance in comparison with the intralayer ones from Ref.~\onlinecite{pore95} (dashed lines).}
\end{figure}
\begin{figure}[tbph]
\includegraphics[width=80mm]{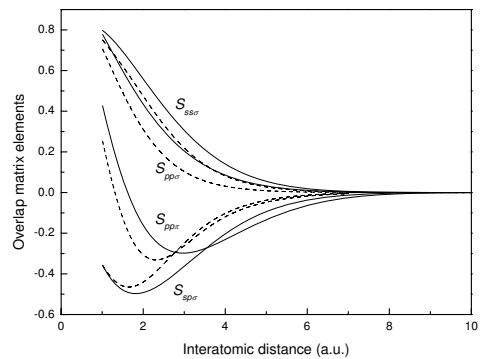} 
\caption{Interlayer overlap matrix elements $S$ (solid lines) as a function of interatomic distance in comparison with the intralayer ones from Ref.~\onlinecite{pore95} (dashed lines). }
\end{figure}

\subsection{The interlayer pair potential}

The interlayer band energy of graphite is an exponentially decreasing function of interlayer separation (Fig. 4). It is therefore necessary to append it  with additional attractive interlayer energy for stabilizing the structure at a finite interlayer separation. The latter energy can be attributed mainly to the weak vdW interactions between the graphene layers. Similarly to the intralayer case, the additional energy is presented as the sum of pair potentials. In the simplest case, the vdW interaction can be described by the power law $-C_6/r^6$, where $r$ is the interatomic separation. A more sophisticated study of the interaction potential in first- and second-order quantum-mechanical perturbation theory yields the series expansion\cite{avoi80,tang84} 

\begin{equation}
V(r)=-\sum_{n=3}^{\infty}C_{2n}r^{-2n}\label{a4}
\end{equation}
where the various power terms account for dispersion and multipole interactions. 

In empirical models, the description of the vdW interactions is usually done by retaining the first term in Eq. 1 and fitting the parameter $C_6$ to the experimental interlayer separation and elastic constant $C_{33}$ of graphite.\cite{giri56,lii89} The cohesive energy is predicted in the range $100-200$ meV/atom. A model potential with several parameters has been fitted to the experimental values of the interlayer separation, the elastic constant $C_{33}$, and the cohesive energy $E_c$ of graphite.\cite{chen13} The elastic constant $C_{44}$, describing the shear deformation of graphite parallel to the layers, is neither used for fitting the parameters of the potential, nor is predicted sufficiently well.

The ab-initio calculations within the DFT-LDA normally predict the cohesive energy in the range $20-30$ meV/atom, while the DFT-GGA one is much lower.\cite{hase04} Recently, $E_c=55-70$ meV/atom has been obtained  with modified functionals within DFT.\cite{chen13} In a hybrid approach, the DFT interlayer energy has been appended with attractive energy in the form of the sum of pair potentials of Lennard-Jones type to obtain $E_c\approx 55$ meV/atom, in agreement with the experimental data.\cite{hase04} The cohesive energy from empirical models is a few times higher, while the DFT one is lower than the measured values of $43$ meV/atom, $35$ meV/atom, and $52$ meV/atom in experiments based on heat of wetting\cite{giri56}, collapse of bulbs at the ends of multiwall carbon nanotubes\cite{bene98}, and thermal desorption of polyaromatic hydrocarbons\cite{zach04}, respectively. 

The elastic constants $C_{33}$ and $C_{44}$, predicted by the DFT-LDA ($29.5$ and $4.5$ GPa, resp.) and the DFT-GGA  ($42.2$ and $3.9$ GPa, resp.) are in fair agreement with the experimental ones of $36.5$ and $4.5$ GPa, resp., (Ref.~\onlinecite{moun05}). We note that the two elastic constants are simply related to the BM and SM.\cite{kitt05} Indeed, the interacting layers of graphite can be considered as a linear atomic chain with phonon dispersion $\omega=\omega_0\sin\left(qd/2 \right)$, where $\omega_0$ is the BM or SM, $q$ is the wavevector, and $d$ is the interlayer separation. On the other hand, the sound velocity for the breathing and shearing phonon branches is $\omega/q = (1/2\pi) \left(C/\rho \right)^{1/2}$, where $C=C_{33}$ or $C_{44}$ and $\rho$ is the density. The resulting relation $C=\rho \left(\pi\omega_0d \right)^2$ allows using the calculated SM and BM instead of the elastic constants $C_{33}$ and $C_{44}$ for comparison with experiment. The DFT predictions for the SM and BM (e.g., Ref.~\onlinecite{wirt04,moun05}) agree fairly well with the measured ones of $49$ cm$^{-1}$ and $126$ cm$^{-1}$, resp. (Ref.~\onlinecite{nick72}).  

\begin{figure}[tbph]
\includegraphics[width=80mm]{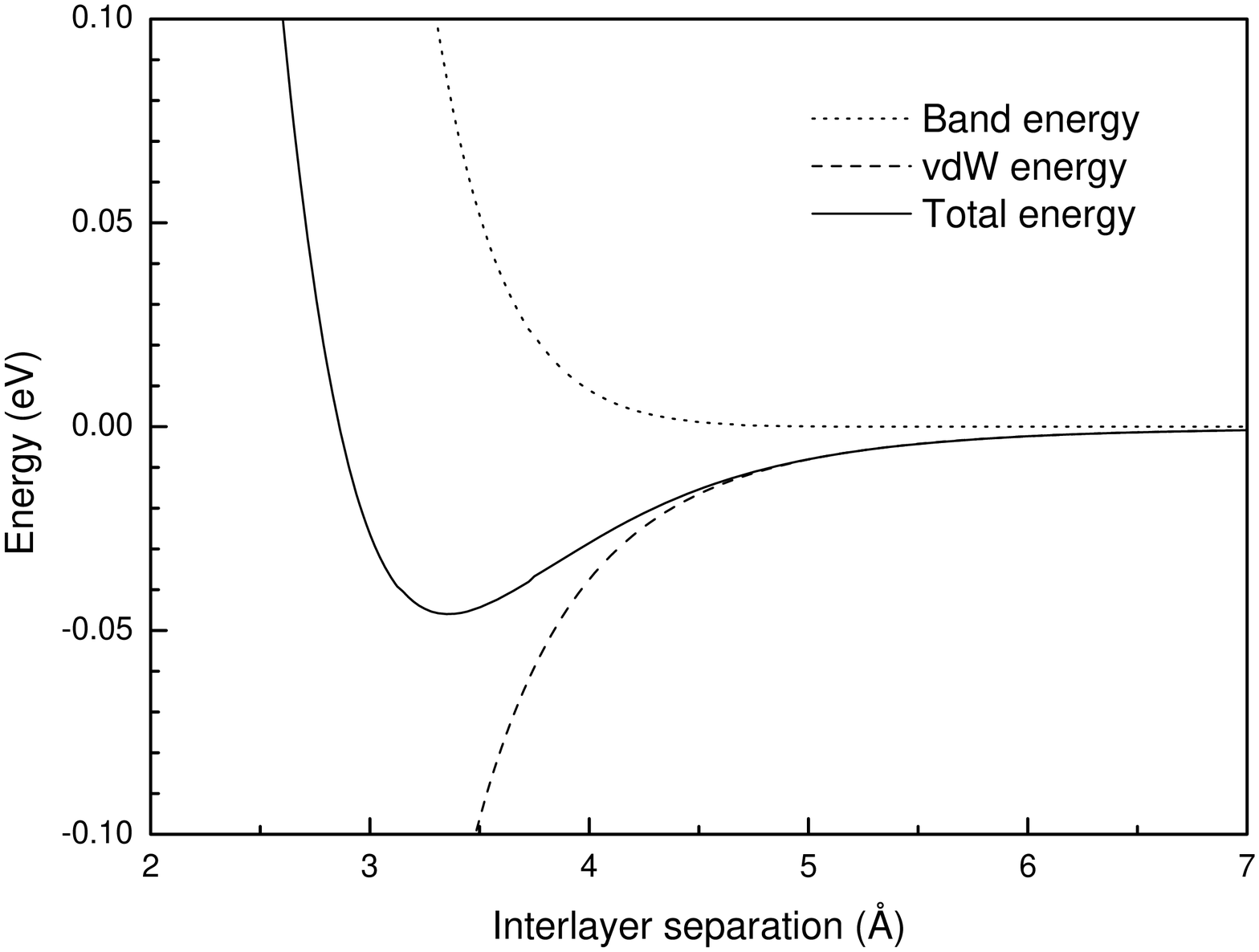} 
\caption{Calculated interlayer interaction energy of graphite (per atom) as a function of interlayer separation. The band energy is drawn with dotted line and the vdW energy is drawn with dashed line. The cohesive energy $E_c = 46$ meV/atom is the minimum of the curve at interlayer separation of $3.35$ \AA. }
\end{figure}

\begin{table}
\caption{\label{tab:table2}
Coefficients $C_{2n}$ (in eV\AA$^{2n}$) obtained by fitting of Eq. 1 to the experimental interlayer separation, SM and BM frequencies (in cm$^{-1}$).  The coefficients at constrained interlayer separation are given in the first three lines, while the last line contains the coefficients at constrained interlayer separation, SM and BM. The calculated cohesive energy $E_c$ (in meV/atom) is shown in the last column.
}
\begin{ruledtabular}
\begin{tabular}{rrrrrr}
$C_6$&$C_8$&$C_{10}$&SM&BM&$E_c$\\
\hline
$49$ & $0$ & $0$ &  $56$&$204$&$74$\\
$40$ & $105$ & $0$ &  $55$&$197$&$67$\\
$5$ & $140$ & $3990$ &  $53$&$137$&54\\
$0$ & $188$ & $4110$ & $49$& $126$ & $46$\\
\end{tabular}
\end{ruledtabular}

\end{table}

The straightforward determination of the parameters of the attractive interlayer potential  can be done by fitting the tight-binding total energy to the ab-initio one. We did not follow this route because the ab-initio approach does not predict the interlayer separation and cohesive energy with sufficient accuracy. Instead, we fitted the parameters of the potential to the experimental values of the interlayer separation, SM and BM of graphite. For every set of derived coefficients, the calculated cohesive energy was checked for agreement with the experimental one. The use of up to three coefficients at constrained interlayer separation yields higher frequencies and cohesive energy that show a trend of decrease with increasing the number of coefficients, as shown in Table II. This behavior persists for up to two coefficients even on imposing constraints on the SM and BM. However, three coefficients could be fitted to reproduce the measured interlayer separation, SM and BM, while the cohesive energy falls in the interval of the measured values of $35-52$ meV/atom. The derived coefficients are $C_{6} = 0 $ eV\AA$^{6}$, $C_{8} = 188$ eV\AA$^{8}$, and $C_{10} = 4110$ eV\AA$^{10}$. The vanishing of $C_6$ may be due to the adopted computational scheme for the tight-binding matrix elements, which utilizes compressed electron density for the carbon atoms. It cannot be excluded that this feature is an intrinsic one, inherited from DFT, which normally yields exponential attraction energy with commonly used exchange-correlation functionals\cite{scha92} or higher-power attraction energy by means of specially designed functionals.\cite{chen13} Equation 1 can be fitted well with more than three coefficients at the chosen constraints but the fit is no longer unique. The calculated vdW energy and the total interlayer energy of graphite are shown as a function of the interlayer separation in Fig. 4.

We apply the interlayer matrix elements and pair potential to the calculation of the phonon dispersion of graphite. The obtained low-frequency dispersion along two high-symmetry directions in the Brillouin zone is shown in Fig. 5. The phonon branches along the  $\Gamma$A direction (wavevector perpendicular to the graphene layers) are acoustic (A) and optical (O) with in-phase and counter-phase movement of the adjacent layers, respectively. These modes can be longitudinal (L) or transverse (T) with atomic displacements along or perpendicular to the wavevector, respectively. The low-frequency phonon branches along the $\Gamma$K direction (wavevector parallel to the layers) can be classified similarly, the only difference being for the transverse branches, which are denoted by T and Z for in-plane and out-of-plane atomic displacements, respectively. The acronyms for the branches along the $\Gamma$A direction are primed. It is clear that, at the $\Gamma$ point, the two components of the doubly-degenerate TO$^{'}$ phonon coincide with the LO and TO phonons. These phonons have shearing layer displacement and are the SM of graphite. Similarly, at the same point, the LO$^{'}$ and ZO phonons are identical. They have breathing layer displacement and are the BM of graphite. The branches with frequency above $\approx 200$ cm$^{-1}$ are almost unchanged with respect to those of graphene\cite{popo10} apart from a splitting up to a few cm$^{-1}$. The shearing and breathing phonon branches along the $\Gamma A$ direction agree well with the experimental ones measured by neutron diffraction\cite{nick72} and x-ray diffraction.\cite{mohr07}
 
\begin{figure}[tbph]
\includegraphics[width=80mm]{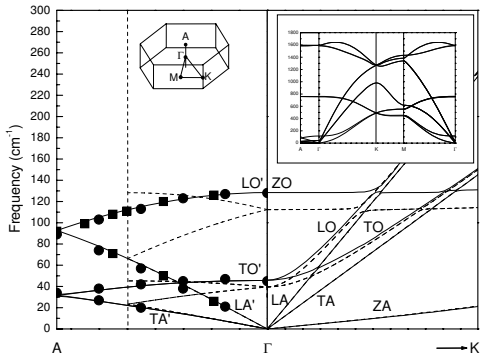} 
\caption{Low-frequency phonon dispersion of Bernal graphite along two high-symmetry directions in comparison with available experimental data, shown by circles\cite{nick72} and squares.\cite{mohr07} The dashed lines are the phonon dispersion of Rhombohedral graphite and the vertical dashed line is the Brillouin zone boundary for this structure. The SM and BM are the doubly-degenerate and non-degenerate optical phonons at the $\Gamma$ point, respectively. Left inset: Brillouin zone of graphite with selected special points. Right inset: Phonon dispersion of Bernal graphite along high-symmetry directions. In the case of graphene, the comparison of the phonon dispersion with experimental data is discussed in Ref.~\onlinecite{popo10}. }
\end{figure}

\subsection{Cohesive energy of FLG}

We use the interlayer matrix elements and pair potential for calculation of static and dynamic properties of several infinite and finite structures with AB and ABC stacking (Fig. 1). The obtained average  interlayer separation for all studied structures is $3.35$\AA~ with deviations within $0.01$ \AA~ for the different structures. The reason for such small deviations can be perceived in the short-range nature of the interlayer interactions, which are negligible beyond the nearest layer. 

The cohesive energies per atom of the infinite AB and ABC structures are almost equal, but the former  has cohesive energy by a few meV larger than the latter. This result allows concluding that the AB crystal is more stable than the ABC one, which explains the predominant observation of three-dimensional graphite with Bernal stacking.  The cohesive energy for several structures with different number of layers is shown in Fig. 6. As a general trend, the cohesive energy increases with increasing the number of layers, which is due to the increase of the average number of neighboring layers from one for 2AB to two for graphite. The cohesive energy of the 2AB structure is almost half of that for Bernal graphite. The difference between the cohesive energies of AB and ABC stacked structures with the same number of layers is negligibly small. 

\begin{figure}[tbph]
\includegraphics[width=80mm]{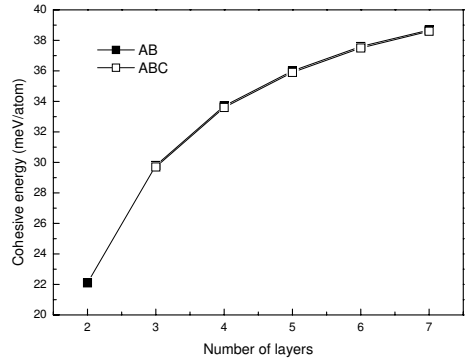} 
\caption{Cohesive energy of several AB and ABC structures as a function of the number of layers. The cohesive energy of Bernal graphite of $46$ meV/atom is about twice that of the 2AB structure.}
\end{figure}

\subsection{Low-frequency phonons of FLG}

The obtained  low-frequency phonon dispersion of three AB stacked FLG is shown in Fig. 7.  As in the case of graphite, there are three acoustic branches of types LA, TA, and ZA. However, in FLG there are $3N-3$ optical branches, among which there $N-1$ ZO branches with atomic dispalcements perpendicular to the layers, as well as $N-1$ TO branches and $N-1$ LO branches with atomic displacements parallel to the layers. At the $\Gamma$ point, the ZO phonons are the BMs and the doubly degenerate LO-TO phonons are the SMs of the FLG. For all optical phonons the adjacent layers show a variety of in-phase and counter-phase displacement. The phonon dispersion of NABC is indistinguishable from that of NAB for the same $N$ (not shown). The derived SMs and BMs for several FLG structures are shown in Fig. 8.

\begin{figure}[tbph]
\includegraphics[width=80mm]{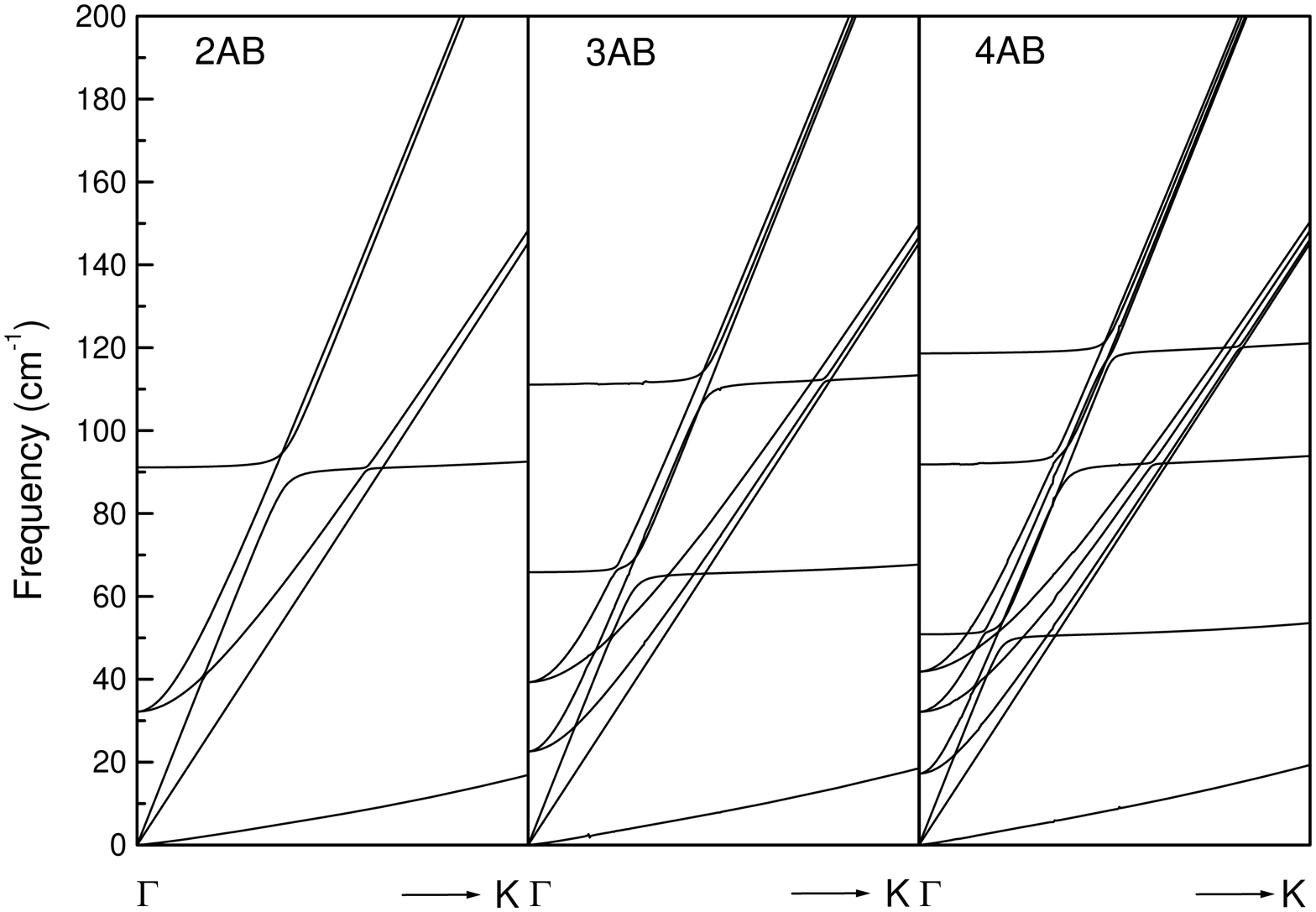} 
\caption{Low-frequency phonon dispersion of three AB structures along the $\Gamma$K direction  as a function of the number of layers. Similarly to the case of graphite, the SMs and BMs are the doubly-degenerate and non-degenerate optical phonons at the $\Gamma$ point, respectively.}
\end{figure}

The so-far reported theoretical SMs and BMs of FLG and graphite have been derived within empirical and ab-initio approaches. The use of a generalized force-constant model\cite{jian08} has yielded the value of $106$ cm$^{-1}$ for the BM of 2AB and $150$ cm$^{-1}$ for Bernal graphite. A force-constant dynamical model with parameters fitted to the experimental SM and BM of graphite, has allowed the fair prediction of the SMs and BMs of NAB.\cite{mich08} Ab-initio calculations of the two modes have been performed within the DFT-LDA and the dynamical matrix has been calculated within the DFPT.\cite{saha08,yan08,tan12} This approach predicts the interlayer separation surprisingly well, which can be fortuitous.\cite{wirt04} The utilization of a more sophisticated approximation as the GGA or a specially devised functional to describe the vdW interactions\cite{tan12} within DFT has resulted in overestimation of the interlayer separation. The predicted SM and BM for graphite of $54$ cm$^{-1}$ and $112$ cm$^{-1}$ within the DFT are in fair agreement with the experimental ones.\cite{saha08}. In the latter paper, the calculations of the two modes have been extended to NAB with $N=2-7$. In another DFPT study,\cite{yan08} the calculated SM and BM of graphite were found to underestimate the measured ones by a few per cent.  A modified functional has been used within the DFT-LDA\cite{thon07} to describe the vdW interaction between the layers.\cite{tan12}  The predicted SM for graphite of $44$ cm$^{-1}$ is close to the experimental value, but the BM of $116$ cm$^{-1}$ is by several per cent lower than the measured one. The calculated SMs for NAB with $N=2-5$ have been found to be in very good agreement with the observed ones.

Only SMs and BMs of NAB have so far been observed experimentally. The SMs of NAB with $N=2-11$ have been measured by Raman spectroscopy, where the observation of lines above $10$ cm$^{-1}$ has been made possible by using three BragGrate notch filters in combination with a single monochromator.\cite{tan12} In another work, using a femtosecond pump-probe technique, the SMs of NAB with $N=2-13$ have been detected by the induced changes in the reflectivity.\cite{bosc13} The BMs of NAB with $N=2-20$ have been determined from the two-phonon Raman spectra.\cite{lui13} Our predicted SMs and BMs match well the available experimental data, as shown in Fig. 8.

\begin{figure}[tbph]
\includegraphics[width=80mm]{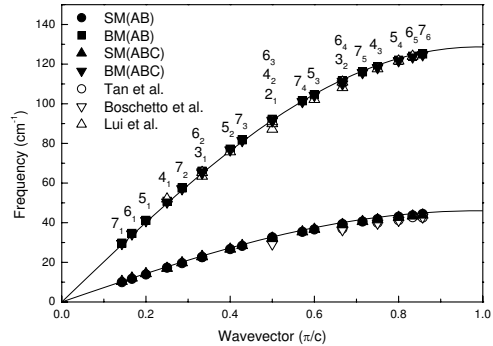} 
\caption{Calculated SMs and BMs for NAB with $N=2-7$ and NABC with $N=3-7$ in comparison with available experimental data for NAB from Tan et al.,\cite{tan12} Boschetto et al.,\cite{bosc13} and Lui et al.\cite{lui13} The position of the $n$-th mode of NAB and NABC is marked by $N_n$. The continuous lines are graphs of Eq. 2 with $n/N$ replaced by a continuous dimensionless wavevector. }
\end{figure}

The interlayer phonon frequencies can be derived by modeling FLG with both AB and ABC stacking as a linear $N$-atom molecule of equally separated identical atoms with nearest-neighbor interactions\cite{kitt05} (see also Ref.~\onlinecite{tan12,lui13}). It is straightforward to obtain the vibrational frequencies of the molecule as
\begin{equation}
\omega_{Nn}=\omega_0\sin(n\pi/2N)
\end{equation}
where the index $n$ enumerates the vibrational modes, $n=1,2,...,N-1$, and $\omega_0$ is the optical phonon frequency of the infinite atomic chain with two-atom unit cell. The latter atomic chain is a model of graphite and $\omega_0$ is the SM or BM of graphite. The calculated frequencies from Eq. 2 correspond almost perfectly to those obtained from the tight-binding calculations and the experimental data, as seen from Fig. 8. The reason for this agreement can be seen in the short-range nature of the interlayer interactions, which do not extend noticeably beyond the nearest layer. 

The presented tight-binding model can be extended to the study of the first- and second-order resonant Raman scattering of FLG. The results of such calculations will be published elsewhere.

\section{Conclusions}

We have demonstrated that the DFT-derived tight-binding parameters can  successfully be applied to the derivation of the phonon dispersion of FLG and graphite. The obtained low-frequency phonons of these structures agree well with the available experimental data. The  interlayer model parameters together with the previously reported intralayer ones can be used for static and dynamic calculations of other layered carbon systems. 

\acknowledgments

V.N.P. acknowledges financial support from Project 316309: INERA - $^{''}$Research and Innovation Capacity Strengthening of ISSP-BAS in Multifunctional Nanostructures$^{''}$.

%

\end{document}